\documentclass[onecolumn,preprintnumbers,elsart]{revtex4}
\usepackage{amsmath}

\usepackage{mathrsfs}
\usepackage{graphicx}
\usepackage{dcolumn}
\usepackage{bm}
\usepackage[center]{subfigure}

\begin{document}

\title{Discrete solitons and vortices on two-dimensional lattices of $%
\mathcal{PT}$-symmetric couplers}
\author {Zhaopin Chen$^{1}$}
\author{Jingfeng Liu$^{1}$}
\author{Shenhe Fu$^{2}$}
\author{Yongyao Li$^{1}$}
\email{yongyaoli@gmail.com}
\author{Boris A. Malomed$^{3}$}

\affiliation{$^{1}$Department of Applied Physics, South China
Agricultural University, Guangzhou 510642, China \\
$^{2}$State Key Laboratory of Optoelectronic Materials and Technologies,\\
Sun Yat-sen University, Guangzhou 510275, China\\
$^{3}$ Department of Physical Electronics, School of Electrical
Engineering, Faculty of Engineering, Tel Aviv University, Tel Aviv
69978, Israel.}

\begin{abstract}
We introduce a 2D network built of $\mathcal{PT}$-symmetric dimers with
on-site cubic nonlinearity, the gain and loss elements of the dimers being
linked by parallel square-shaped lattices. The system may be realized as a
set of $\mathcal{PT}$-symmetric dual-core waveguides embedded into a
photonic crystal. The system supports $\mathcal{PT}$-symmetric and
antisymmetric fundamental solitons (FSs) and on-site-centered solitary
vortices (OnVs). Stability of these discrete solitons is the central topic
of the consideration. Their stability regions in the underlying parameter
space are identified through the computation of stability eigenvalues, and
verified by direct simulations. Symmetric FSs represent the system's ground
state, being stable at lowest values of the power, while anti-symmetric FSs
and OnVs are stable at higher powers. Symmetric OnVs, which are also stable
at lower powers, are remarkably robust modes: on the contrary to other $%
\mathcal{PT}$-symmetric states, unstable OnVs do not blow up, but
spontaneously rebuild themselves into stable FSs.
\end{abstract}

\maketitle






\section{Introduction}

Dynamics of light fields in dissipative media keeps drawing much interest in
nonlinear optics \cite{review0,review}. A special kind of such systems with
exactly balanced spatially separated gain and loss realize the general
concept of the parity-time ($\mathcal{PT}$)-symmetry \cite{Bender1,Bender2}.
$\mathcal{PT}$-symmetric systems are represented by Hamiltonians containing
a complex potential, $V(\mathbf{r})$, which is subject to the symmetry constraint, $V(%
\mathbf{r})=V^{\ast }(-\mathbf{r})$. In optics this condition can be
realized by guiding light through a medium with a complex refractive index
satisfying condition $n(\mathbf{r})=n^{\ast }(-\mathbf{r})$ \cite{Muga,Makris,Klaiman,Guo,Longhi1,Longhi2,Makris2,coupled-mode}, hence its real and imaginary part, the latter one
representing the combination of spatially distributed gain and loss, must
be, respectively, even and odd functions of spatial coordinates.

Unlike generic dissipative systems, in which the balance between gain and
loss selects parameters of isolated stable modes (\textit{attractors}), $%
\mathcal{PT}$-symmetric settings support continuous families of modes,
sharing this property with conservative media, which was a motivation for
many experimental \cite{Regensburger,exper} and theoretical studies,
including the consideration of a great diversity of nonlinear $\mathcal{PT}$%
-symmetric systems \cite{Musslimani,Zhuxing,Yaro,Lihuagang,Radik,oligo,Miro,Barash,Nixon,Abd,Alexeeva,Uwe,birefr,Thawatchai,Flach,Burlak,Hadi,Romania,restoration,Yang,SYS,unbreakable}. In the latter case,
solitons are a major subject of the studies.

An important variety of $\mathcal{PT}$-symmetric systems in optics are
discrete ones. The simplest among them, which was actually the first $%
\mathcal{PT}$-symmetric system created experimentally \cite{Moti}, is
realized as a pair of linearly coupled waveguides carrying gain and loss
with equal strengths (a \textit{dimer }\cite%
{oligo,Miro,Uwe,birefr,Flach,Hadi}). An extension to various more complex
discrete systems is provided by arrays of linearly coupled dimers \cite{Dmitriev,Suchkov,Suchkov2,Suchkov3,Zezyulin,Konotop,Leykam,Tsironis,Peli,Xiangyu,AubryAndre}. In particular, robust one-dimensional (1D)
discrete solitons \cite{Suchkov} and scattering states \cite{Suchkov3} are
supported by a chain of dimers with the cubic nonlinearity, in which active
(gain-carrying) and passive (lossy) elements are coupled horizontally, while
at each site the active and passive poles are coupled vertically.
Accordingly, the chain is described by a system of linearly coupled discrete
nonlinear Schr\"{o}dinger (DNLS) equations with equal strengths of linear
gain and loss in the two components. Stability, mobility and interactions
between fundamental discrete solitons in this system were studied by means
of numerical methods and analytical approximations.

There are fewer publications which reported $\mathcal{PT}$-symmetric
solitons in 2D settings \cite{Nixon,Burlak,SYS}. The objective of the
present work is to introduce a 2D network of horizontally coupled nonlinear
dimers, which is a generalization of the 1D system from \cite{Suchkov}, and
can also be realized in optics, using a photonic-crystal matrix into which
nonlinear dual-core waveguides with the balanced gain and loss are embedded
(see Fig. \ref{Model} below). We construct families of fundamental and
vortical two-component discrete solitons in this 2D system and analyze their
stability.

The paper is structured as follows. The model is introduced in Section 2.
Fundamental and vortex solitons are produced in Sections 3 and 4,
respectively. The paper is concluded by Section 5.

\section{Model}

\begin{figure}[tbp]
\centering{\label{fig1a} \includegraphics[scale=0.4]{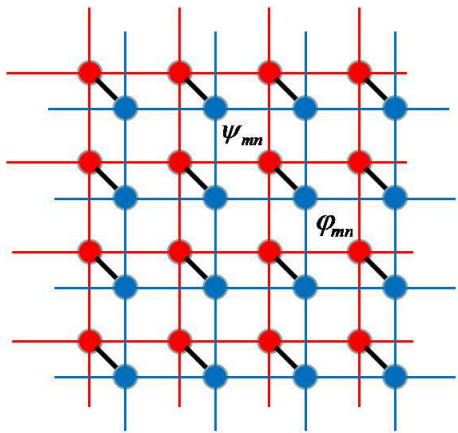}}
\caption{The transverse cross-section of the waveguiding system built as a
juxtaposition of networks of active (red) and passive (blue) elements, which
are coupled into $\mathcal{PT}$-symmetric dimers (the couplings are
designated by short diagonal segments). The elements are connected into the
networks by thin plates forming the double photonic crystal. The propagation
axis ($z$) is directed perpendicular to the plane of the drawing.}
\label{Model}
\end{figure}

The network model, outlined above as the 2D extension of the chain of dimers
introduced in \cite{Suchkov}, is based on a system of\ two linearly coupled
normalized DNLS equations for amplitudes $\psi _{m,n}(z)$ and $\varphi
_{m,n}(z)$ of the electromagnetic field propagating in the active and
passive cores of dual-core waveguides (that represent the $\mathcal{PT}$%
-symmetric dimers), which, as schematically shown in Fig. \ref{Model}, are
embedded into a quasi-2D photonic crystal, built of thin plates connecting
the active and passive cores into respective sub-networks:
\begin{eqnarray}
&&i{\frac{d}{dz}}\psi _{m,n}=-{\frac{C}{2}}(\psi _{m,n+1}+\psi _{m,n-1}+\psi
_{m-1,n}+\psi _{m+1,n}-4\psi _{m,n})-|\psi _{m,n}|^{2}\psi _{m,n}-\varphi
_{m,n}+i\gamma \psi _{m,n},  \notag   \\
&&i{\frac{d}{dz}}\varphi _{m,n}=-{\frac{C}{2}}(\varphi _{m,n+1}+\varphi
_{m,n-1}+\varphi _{m-1,n}+\varphi _{m+1,n}-4\varphi _{m,n})-|\varphi
_{m,n}|^{2}\varphi _{m,n}-\psi _{m,n}-i\gamma \varphi _{m,n}. \label{basicEq}
\end{eqnarray}%
Here $\left( m,n\right) $ are discrete coordinates in the transverse plane, $%
z$ is the propagation distance, the coefficient of the coupling between the
active and passive cores is scaled to be $1$, $\gamma >0$ is the gain-loss
coefficient, and $C$ is coupling constant in the networks. Localized states
produced by Eq. (\ref{basicEq}) are characterized by the total power,
\begin{equation}
P=\sum {}_{m,n}\left( |\psi _{m,n}|^{2}+|\varphi _{m,n}|^{2}\right) .
\label{P}
\end{equation}

To estimate physical parameters of the setting, we assume that it is built of
AlGaAs (the nonlinear coefficient of this material is $n_{2}\simeq 1.5\times
10^{-13}$ cm$^{2}$/W \cite{Aitchison}), with core area $4$ $\mathrm{\mu }$m$^{2}$
of the lattice site, and refractive indices of the core and thin plates $%
3.263$ and $3.256$, respectively (the corresponding refractive-index
difference can be produced by varying the share of aluminium in the
composition of those parts). The standard coupled-mode theory \cite%
{Yariv,Zhijie} shows that, if we fix the distance between the adjacent
active and passive lattice sites as $4$ $\mathrm{\mu }$m, the coupling
coefficient between them is $3.6$ mm$^{-1}$. Because we scale
this coefficient to be $1$ in Eq. (\ref{basicEq}), we can estimate the
accordingly normalized values of the coupling and gain-loss coefficients, $C$
and $\gamma $, and the unit of the total power, $P_{0}=0.55$ kW. For
example, the lattice spacing$\ 8$ $\mathrm{\mu }$m and the physical value of
the gain-loss coefficient, $1.8$ mm $^{-1}$, imply $C=0.3$ and $\gamma =0.5$%
. These values are realistic for the experimental realization, and, on the other hand,
their normalized values give rise to nontrivial results, as shown below.

Stationary modes with real propagation constant $k$ are looked for as
solutions to Eq. (\ref{basicEq}) as usual, $\left\{ \psi _{m,n}(z),\varphi
_{m,n}(z)\right\} =e^{ikz}\left\{ u_{m,n},v_{m,n}\right\} $, where
stationary amplitudes obey equations in the form of%
\begin{eqnarray}
\left( k+i\gamma \right) u_{m,n} &=&{\frac{C}{2}}%
(u_{m,n+1}+u_{m,n-1}+u_{m-1,n}+u_{m+1,n}-4u_{m,n})+|u_{m,n}|^{2}u_{m,n}+v_{m,n},
\notag \\
\left( k-i\gamma \right) v_{m,n} &=&{\frac{C}{2}}%
(v_{m,n+1}+v_{m,n-1}+v_{m-1,n}+v_{m+1,n}-4v_{m,n})+|v_{m,n}|^{2}v_{m,n}+u_{m,n}.
\label{UV}
\end{eqnarray}%
Because our study is focused on localized modes, we apply zero boundary
conditions at infinity (in fact, at edges of the numerical domain).

In the numerical simulations, it is convenient to combine $u_{m,n}$ and $%
v_{m,n}$ into a single \textquotedblleft long" vector, $U$, of length $2N^{2}
$, where $N\times N$ is the size of the numerical domain. Thus, Eq. (\ref{UV}%
) is rewritten in the vectorial form:
\begin{equation}
\hat{L}U-\mathrm{\mathbf{diag}}(|U|^{2})U=-\left( k+i\hat{\Gamma}\right) U,
\label{basicEq2}
\end{equation}%
where $\mathrm{\mathbf{diag}}(|U|^{2})$ means a diagonal matrix with the
respective elements, the other matrices being $\hat{L}=\left(
\begin{array}{cc}
\mathbf{D} & -\mathbf{I} \\
-\mathbf{I} & \mathbf{D}%
\end{array}%
\right) $ and $\hat{\Gamma}=\gamma \left(
\begin{array}{cc}
\mathbf{I} & 0 \\
0 & -\mathbf{I}%
\end{array}%
\right) $, where $\mathbf{I}$ and $\mathbf{D}$\ are, respectively, the $%
N^{2}\times N^{2}$ unit and linear-coupling matrices, the latter one built
of elements $D_{j,j^{\prime }}=-\left( C/2\right) (\delta _{j,j^{\prime
}-N}+\delta _{j,j^{\prime }-1}+\delta _{j,j^{\prime }+1}+\delta
_{j,j^{\prime }+N}-4\delta _{j,j^{\prime }})$. The stability of the
stationary modes is investigated in a numerical form, computing eigenvalues
for small perturbations and verifying the results by direct simulations.
Linearized equations for perturbation eigenmodes are produced by writing the
perturbed solution, in terms of the \textquotedblleft long vector", $U$, as $%
\left\{ \psi _{m,n}(z),\varphi _{m,n}(z)\right\} =e^{ikz}(U+\alpha
e^{-i\lambda z}+\beta ^{\ast }e^{i\lambda ^{\ast }z})$, where $\alpha $ and $%
\beta $ are the respective vectors combining amplitudes of the eigenmodes,
and $\ast $ stands for the complex conjugation:

\begin{equation}
\left(
\begin{array}{cc}
\hat{L}+k-2\mathrm{\mathbf{diag}}(|U|^{2})+i\hat{\Gamma} & -\mathrm{\mathbf{%
diag}}(U^{2}) \\
\mathrm{\mathbf{diag}}(U^{\ast 2}) & -\hat{L}-k+2\mathrm{\mathbf{diag}}%
(|U|^{2})+i\hat{\Gamma}%
\end{array}%
\right) \left(
\begin{array}{c}
\alpha \\
\beta%
\end{array}%
\right) =\lambda \left(
\begin{array}{c}
\alpha \\
\beta%
\end{array}%
\right) .  \label{eigenvalue}
\end{equation}%
The underlying solution, $U$, is stable if all eigenvalues $\lambda $ are
real.

\section{Fundamental solitons}

\subsection{Symmetric modes}

As suggested by the analysis of the $\mathcal{PT}$-symmetric coupler \cite%
{Radik,Barash}, the parity-time symmetry of Eqs. (\ref{basicEq}) remains
unbroken at $\gamma <1$, in which case the system of two coupled equations (%
\ref{UV})\ can be reduced to a single one,%
\begin{equation}
\left( -k\pm \sqrt{1-\gamma ^{2}}\right) u_{m,n}+{\frac{C}{2}}\left(
u_{m,n+1}+u_{m,n-1}+u_{m-1,n}+u_{m+1,n}-4u_{m,n}\right)
+|u_{m,n}|^{2}u_{m,n}=0,  \label{U}
\end{equation}%
by substitution%
\begin{equation}
v_{m,n}=e^{i\chi _{\pm }}u_{m,n},~\chi _{+}=\arcsin \gamma ,~~\chi _{-}=\pi
-\arcsin \gamma ,  \label{chi}
\end{equation}%
with $\pm $ in Eq. (\ref{U}) corresponding to $\chi _{\pm }$ in Eq. (\ref%
{chi}). Because in the limit of $\gamma =0$
relation (\ref{chi}) with $\chi _{+}$ and $\chi _{-}$ represents,
respectively, symmetric and antisymmetric states in the two-component
conservative system, these solutions may be naturally named $\mathcal{PT}$%
-symmetric and antisymmetric ones \cite{Radik}.

Discrete 2D fundamental solitons (FSs) in the conservative version of Eqs. (%
\ref{basicEq}) and (\ref{UV}) with $\gamma =0$ were previously studied in
the context of the spontaneous formation of asymmetric solitons, with $%
u_{m,n}\neq v_{m,n}$, from the symmetric ones \cite{Herring}. The transition
to asymmetric solitons is not relevant for $\mathcal{PT}$-symmetric systems,
where the spontaneous symmetry breaking causes blow-up, rather than
emergence of asymmetric modes \cite{Radik,Barash}) (recently, it was
demonstrated that asymmetric solitons may exist in a specially designed 1D $%
\mathcal{PT}$-symmetric model \cite{Yang}). However, the instability may
lead, instead of the blow-up, to transformation into a stable soliton of
another type, if different solitons species coexist in the system. In
particular, it is demonstrated below that unstable $\mathcal{PT}$-symmetric
solitary vortices \emph{do not} blow up, but rather rebuild themselves into
stable symmetric FSs.

Equation (\ref{U}) is the usual stationary form of the 2D DNLS equation,
which gives rise to discrete FSs \cite{Panos} and solitary vortices \cite%
{Malomed} whose total power (\ref{P}) grows with the increase of the
effective propagation constant,
\begin{equation}
k_{\mathrm{eff}}=k\mp \sqrt{1-\gamma ^{2}},  \label{keff}
\end{equation}%
so that these modes satisfy the Vakhitov-Kolokolov criterion \cite{VK,Berge'}%
, $dP/dk>0$, which is a necessary, but not sufficient, condition for their
stability. Therefore, for given $k$, term $\mp \sqrt{1-\gamma ^{2}}$ in Eq. (%
\ref{keff}) makes the total power of the $\mathcal{PT}$-symmetric FS \emph{%
smaller} than that of its antisymmetric counterpart. The latter fact, in
turn, implies that the symmetric FS may represent the system's ground state
(it is demonstrated below that this conjecture is true), and it should be
more relevant to the experiment, as the creation of the mode with a smaller
power is, obviously, easier.

\begin{figure}[tbp]
\centering{\label{fig2a} \includegraphics[scale=0.42]{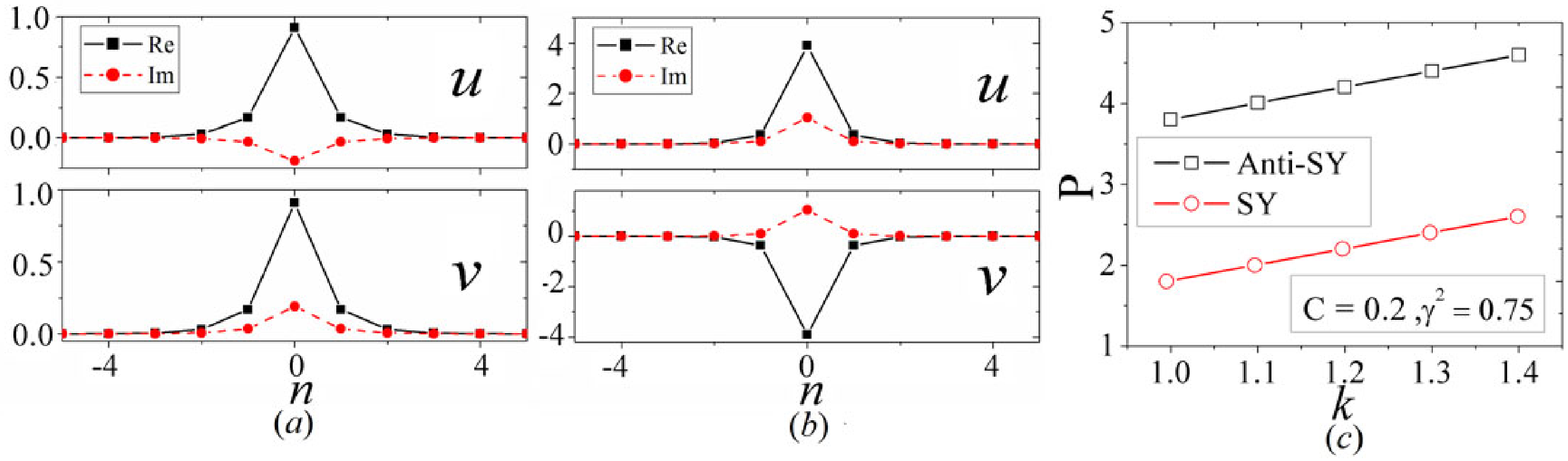}}
\caption{(a) The cross-section, along $m=0$, of a typical $\mathcal{PT}$%
-symmetric fundamental soliton produced by Eq. (\protect\ref{basicEq}), with
$(C,\protect\gamma )=(0.3,0.4)$, and total power $P=2$. Its real and
imaginary parts precisely obey Eq. (\protect\ref{chi}) with $\protect\chi =%
\protect\chi _{+}$. Both the computation of stability eigenvalues and direct
simulations demonstrate that this soliton is stable. (b) The cross-section
(along $m=0$) of a typical stable $\mathcal{PT}$-antisymmetric fundamental
soliton, which is also stable, produced by stationary equations (\protect\ref%
{basicEq}) for $(C,\protect\gamma )=(3,0.5)$, and $k=10$. The total power of
this soliton is $P=33.8$. (c) $P(k)$ dependences for the symmetric
(\textquotedblleft SY") and antisymmetric (\textquotedblleft Anti-SY")
fundamental solitons, with $C=0.2$ and $\protect\gamma =\protect\sqrt{3}/2$.
The fitting result for them is $P(k)=4C+2k\mp 1$, which is consistent with
Eq. (\protect\ref{P(k)}).}
\label{ExpFS}
\end{figure}

Stationary solutions for symmetric FSs were produced, first, by means of the
imaginary-time-propagation method (ITM) \cite{gs}, applied to Eq. (\ref%
{basicEq}). For these computations, the set of control parameters includes
the gain-loss and coupling strengths, $\left( \gamma ,C\right) $, and the
total power of the soliton, $P$. It was checked that precisely the same
family of symmetric solutions is generated by the Newton's method applied to
stationary equations (\ref{basicEq2}). The fact that these solutions are
produced by the ITM method strongly suggests that they indeed represent the
system's ground state \cite{Tosi}. It is also relevant to stress that Eq. (%
\ref{basicEq2}) produces exactly the same solutions as the reduced equation (%
\ref{U}), although reduction (\ref{chi}) was not applied to Eq. (\ref%
{basicEq2}). A typical example of a stable $\mathcal{PT}$-symmetric FS is
displayed in Fig. \ref{ExpFS}.

For $\gamma =0$, the results corroborate well-known properties of the 2D FSs
in the DNLS equation \cite{Panos}. In particular, symmetric on-site-centered
FSs (their off-site-centered counterparts are completely unstable, as usual
\cite{Panos}, and are not considered here) exist above a power threshold,
\begin{equation}
P>P_{\mathrm{th}}^{(1)}=\mathrm{const}\cdot C,~\mathrm{const}\approx 5.7.
\label{thr1}
\end{equation}%
Because stationary equations (\ref{UV}) for $\mathcal{PT}$-symmetric modes
reduce to the single equation (\ref{U}), it is obvious that $P_{\mathrm{th}%
}^{(1)}$ does not depend on $\gamma $. Further, if the $P(k)$ dependence for
the FS in the single-component DNLS equation with $\gamma =0$ is $P_{1}(k)$,
which is well known in the numerical form \cite{Panos}, then an\emph{\ exact
result} for the two-component $\mathcal{PT}$-symmetric and antisymmetric
FSs, following from Eq. (\ref{U}), is
\begin{equation}
P=2P_{1}\left( k\mp \sqrt{1-\gamma ^{2}}\right) .  \label{exact}
\end{equation}%
Numerical solutions for both types of the solitons completely agree with Eq.
(\ref{exact}) (not shown here in detail). In fact, for both the $\mathcal{PT}
$-symmetric and antisymmetric FSs the $P(k)$ dependence can be rather
accurately fitted to an empiric relation%
\begin{equation}
P_{\mathrm{FS}}(k,\gamma )\approx 2\left( 2C+k\mp \sqrt{1-\gamma ^{2}}%
\right) .  \label{P(k)}
\end{equation}%
Figure \ref{ExpFS}(c) displays the typical examples of such numerical
fitting. In particular, the known results for the stable 2D FS in the usual
DNLS equation, corresponding to $\gamma =0$ \cite{Panos}, are indeed close
to the approximate linear dependence obtained for this case from Eq. (\ref%
{P(k)}), $P_{1}(k)\approx 2C+k$.

An essential finding is that the symmetric FS becomes unstable above a
higher threshold value of the total power, $P=P_{\mathrm{th}}^{(2)}$. This
stability boundary may be understood as the one at which the symmetric
soliton is destabilized by the spontaneous symmetry breaking, modified by
the presence of $\gamma >0$ in comparison with the similar effect in the
conservative system with $\gamma =0$ \cite{Herring}. Because asymmetric
solitons cannot exist in the system with the balanced gain and loss, this
symmetry breaking always leads to blow-up of the FS in the active component [%
$\psi _{m,n}$ in Eq. (\ref{basicEq})]. The blow-up is quite similar to that
shown below in Fig. \ref{instability}(b) for an antisymmetric vortex.

\begin{figure}[tbp]
\centering{\label{fig3a} \includegraphics[scale=0.27]{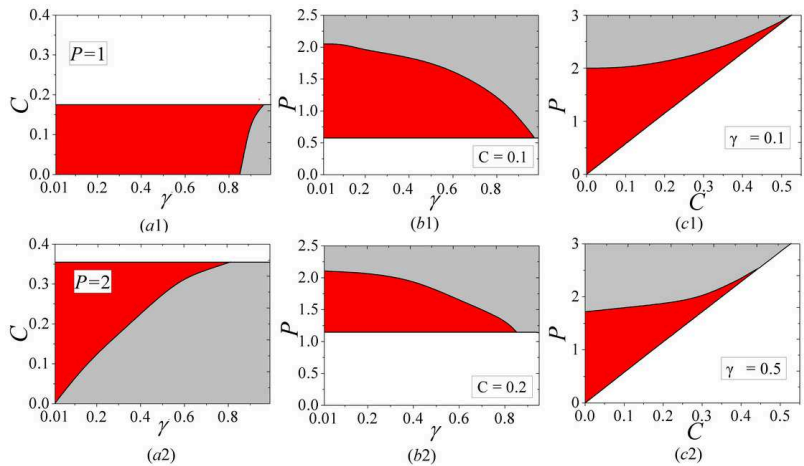}}
\caption{Here and in similar figures below, red and gray colors represent,
respectively, stability and instability areas -- here, for $\mathcal{PT}$%
-symmetric fundamental solitons -- as found from the computation of the
stability eigenvalues [see Eq. (\protect\ref{eigenvalue})], and verified by
direct simulations. In white areas, soliton solutions cannot be found. The
red-white and red-gray boundaries correspond, severally, to threshold values
$P_{\mathrm{th}}^{(1)}$ and $P_{\mathrm{th}}^{(2)}$ (see the text).}
\label{SAFS}
\end{figure}

The most essential results of the numerical analysis of the $\mathcal{PT}$%
-symmetric FSs are summarized in Fig. \ref{SAFS}, in the form of stability
diagrams in parameter planes $\left( \gamma ,C\right) $, $\left( \gamma
,P\right) $, and $\left( C,P\right) $. The horizontal and diagonal cutoffs
in the $\left( \gamma ,C\right) $, $\left( \gamma ,P\right) $ and $\left(
C,P\right) $ planes exactly represent the threshold power, as given by Eq. (%
\ref{thr1}). A natural trend is shrinkage of the stability area with the
increase of $\gamma $, and its full disappearance close to $\gamma =1$. It
is natural as well that the stability area expands with the increase of $C$
at fixed $P$, because the large coupling constant makes the modes,
effectively, \textquotedblleft less nonlinear", for the given total power,
and this, in turn, suppresses the trend to the nonlinearity-driven
spontaneous symmetry breaking which destabilizes the soliton.

\subsection{Antisymmetric solitons}

\begin{figure}[tbp]
\centering{\label{fig4a} \includegraphics[scale=0.35]{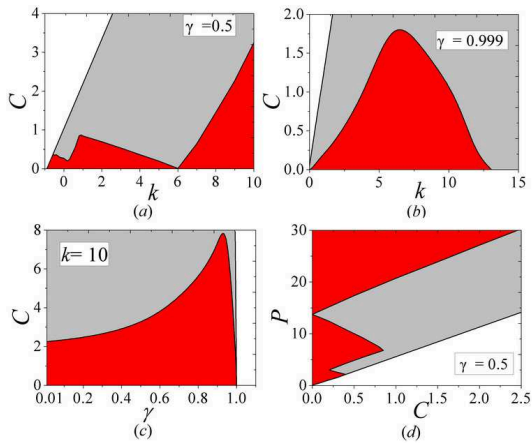}}
\caption{(a,b) The stability area (red) of the antisymmetric fundamental
solitons for (a) $\protect\gamma =0.5$ and (b) $\protect\gamma =0.999$. (c)
The stability area in the $(\protect\gamma ,C)$ plane for $k=10$. (d) The
stability diagram for $\protect\gamma =0.5$ in the $(C,P)$ plane. }
\label{SAFANS}
\end{figure}

It has been found that the system supports stable and unstable antisymmetric
FSs too [see a typical examples of a stable one in Fig. \ref{ExpFS}(b)]. The
antisymmetric FSs cannot be produced by the ITM, which suggests that, as
argued above, they do not represent the system's ground state. Nevertheless,
they have been generated by the Newton's method applied to Eq. (\ref%
{basicEq2}). For this reason, the propagation constant, $k$, rather than the
total power, $P$, plays the role of the control parameter for the
antisymmetric solitons. Their stability was identified through a numerical
solution of Eq. (\ref{eigenvalue}), and verified by means of direct
simulations of Eq. (\ref{basicEq}). Results of the stability analysis are
collected in Fig. \ref{SAFANS}, in which panels (a-c) display the stability
area in the $\left( k,C\right) $ and $\left( \gamma ,C\right) $ panels. In
addition, the stability area is mapped into the $\left( C,P\right) $ plane
in plot \ref{SAFANS}(d), using the $P(k)$ dependence [see Eq. (\ref{P(k)})].
In the latter panel, the boundary of the white non-existence area is given
by Eq. (\ref{thr1}), i.e., it is exactly the same as in Fig. \ref{SAFS}(c1)
and \ref{SAFS}(c2), once the existence of both symmetric and antisymmetric
FSs is determined by the same equation, Eq. (\ref{U}).

We have found that, at $\gamma <0.95$, the stability area shown in panel \ref%
{SAFANS}(d) tends to extend to $P\rightarrow \infty $ [or, in other words,
to $k\rightarrow \infty $ in panel \ref{SAFANS}(a)], up to the accuracy
limits of the numerical calculations. This conclusion is in drastic contrast
with the situation shown for the symmetric FSs in Fig. \ref{SAFS}, where the
stability is always bounded from above, in terms of $P$. The explanation for
the stability of the antisymmetric solitons at large $P$ is straightforward:
unlike their symmetric counterparts, they do not undergo the destabilization
via the symmetry breaking at large $P$. However, at $\gamma >0.95$, the
stability area becomes finite in terms of $k$, see Fig. \ref{SAFANS}(c) for $%
\gamma =0.999$. Although the area is not vanishingly small at $\gamma $ so
close to $1$, it exactly collapses to nil at $\gamma =1$, as it should, as
the $\mathcal{PT}$ symmetry gets completely broken at this point.

A general conclusion about the antisymmetric FSs (which pertains equally
well to antisymmetric vortex solitons, see below) is that, although they
tend to be stable at large values of the power, where their symmetric
counterparts are unstable, they are less suitable for the experimental
realization just because they demand large powers for the stability, while
the symmetric FS, which represents the ground state of the system, may be
stable at lower levels of the power.

\section{Vortex soliton}

To the best of our knowledge, the symmetry breaking of discrete vortex
solitons in 2D two-component linearly-coupled lattices was not investigated
even in the conservative model, with $\gamma =0$. We have found symmetric,
antisymmetric, and asymmetric solitary vortices, and the symmetry-breaking
transition, in the model based on Eqs. (\ref{basicEq}) and (\ref{UV}) with $%
\gamma =0$. These findings will be reported elsewhere, while here we focus
on $\mathcal{PT}$-symmetric and antisymmetric vortices in the system with $%
\gamma >0$ (in fact, it is shown below that, for symmetric vortices, the
stability area only weakly changes with the variation of $\gamma $, hence
the present results give an idea of the stability of symmetric vortices in
the conservative system too). Because the ITM cannot converge to the
definitely non-ground-state vortical modes, the solutions were found by
means of the Newtons's method applied to Eq. (\ref{basicEq2}), hence the
control parameter of the solution families is the propagation constant, $k$,
rather than total power $P$.

It is well known that two types of vortices are possible in discrete
systems, \textit{viz}., on- and off-site-centered ones \cite{Panos} (alias
\textquotedblleft rhombuses" and \textquotedblleft squares"). Both can be
produced by the Newton's method applied to Eq. (\ref{UV}), starting from the
following inputs:

\begin{equation}
\left\{ u_{m,n},v_{m,n}\right\} _{\mathrm{on}}=\left\{
\begin{array}{c}
\left\{ U,V\right\} ~~\mathrm{at}~~\left( m,n\right) =\left( 0,1\right) , \\
i\left\{ U,V\right\} ~~\mathrm{at}~~\left( m,n\right) =\left( 1,0\right) ,
\\
-\left\{ U,V\right\} ~~\mathrm{at}~~\left( m,n\right) =\left( -1,0\right) ,
\\
-i\left\{ U,V\right\} ~~\mathrm{at}~~\left( m,n\right) =\left( 0,-1\right) ,
\\
0,~\mathrm{at~all~others}~~\left( m,n\right) ,~~~%
\end{array}%
\right.  \label{on}
\end{equation}

\begin{equation}
\left\{ u_{m,n},v_{m,n}\right\} _{\mathrm{off}}=\left\{
\begin{array}{c}
\left\{ U,V\right\} ~~\mathrm{at}~~\left( m,n\right) =\left( 0,0\right) , \\
i\left\{ U,V\right\} ~~\mathrm{at}~~\left( m,n\right) =\left( 1,0\right) ,
\\
-\left\{ U,V\right\} ~~\mathrm{at}~~\left( m,n\right) =\left( 1,1\right) ,
\\
-i\left\{ U,V\right\} ~~\mathrm{at}~~\left( m,n\right) =\left( 0,1\right) ,
\\
0,~\mathrm{at~all~others}~~\left( m,n\right) ,~%
\end{array}%
\right.  \label{off}
\end{equation}%
where $U$ and $V$ are real constants. 

In agreement with the known result for the usual 2D\ DNLS equation with $%
\gamma =0$ \cite{Malomed}, we have found that the off-site vortices are
unstable at practically all values $C>0$ (they may be stable at extremely
small $C$), therefore in what follows below we consider only the modes of
the on-site types.

\begin{figure}[tbp]
\centering{\label{fig5a} \includegraphics[scale=0.27]{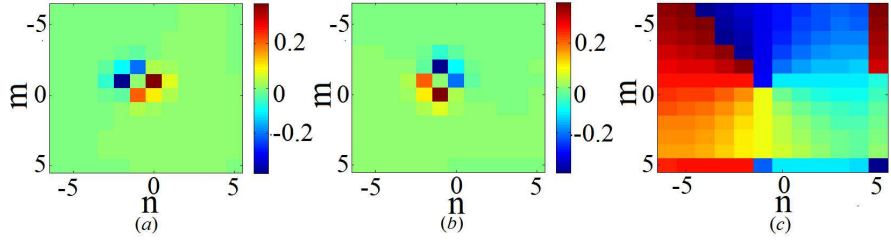}}
\caption{Real (a) and imaginary (b) parts, and the phase structure (c), of field
$u_{m,n}$ of a typical stable $\mathcal{PT}$-symmetric vortex soliton, for $(C,\protect\gamma )=(0.06,0.4)$
and $k=1$ ($P=1.65$).}
\label{ExpPTsyVS}
\end{figure}

\begin{figure}[tbp]
\centering{\label{fig6a} \includegraphics[scale=0.27]{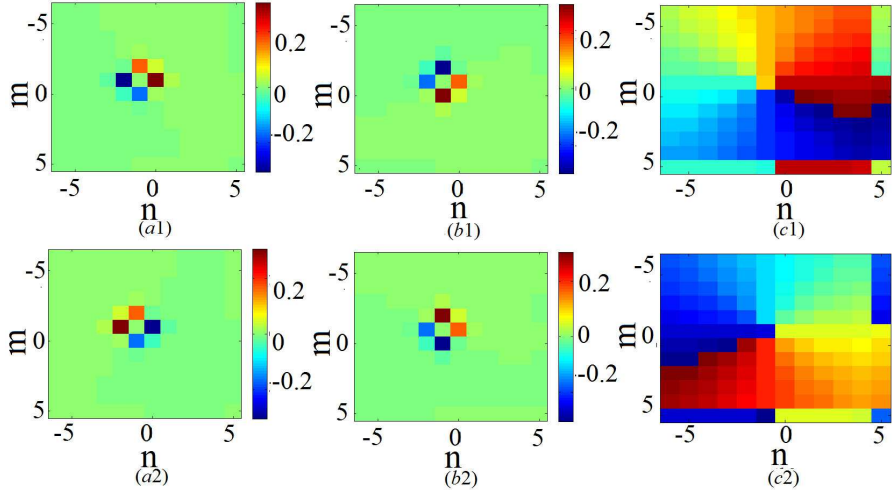}}
\caption{The same as in Fig. \protect\ref{ExpPTsyVS}, but for a stable $%
\mathcal{PT}$-antisymmetric vortex, for $(C,\protect\gamma )=(2,0.85)$ and $%
k=8$ ($P=100.3$). In this figure, both components are
displayed, to stress that relation (\protect\ref{chi}) for the antisymmetric
vortex, with $\protect\chi =\protect\chi _{-}$, gives rise to rotation of
the phase pattern by angle $\protect\pi -\arcsin \left( 0.85\right) \approx
0.7\protect\pi $.}
\label{ExpPtAntiVS}
\end{figure}

As well as in the case of FSs, stationary solutions for on-site-centered
vortices (OnVs) are subject to reduction (\ref{chi}), which leads to the
single equation (\ref{U}), hence the OnV may also be classified into $%
\mathcal{PT}$ symmetric or antisymmetric varieties. Examples of stable
symmetric and antisymmetric ones are displayed in Figs. \ref{ExpPTsyVS} and %
\ref{ExpPtAntiVS}.

\begin{figure}[tbp]
\centering{\label{fig7a} \includegraphics[scale=0.27]{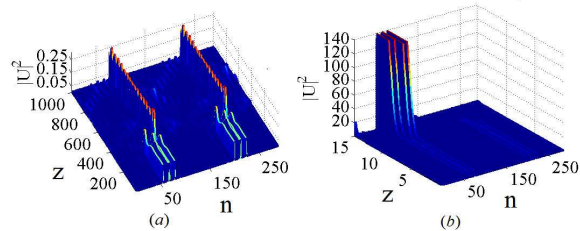}} \centering{\ }
\caption{(a) The instability evolution of a symmetric vortex, with $C=0.06,%
\protect\gamma =0.4$, and $k=0.96$ ($P=1.37$). This unstable vortex\emph{\
does not} blow up, but spontaneously transforms into a stable $\mathcal{PT}$%
-symmetric fundamental soliton. In these panels, the two components of the
vortex are juxtaposed, to display the evolution in cross section $m=0$. (b)
The instability development in an antisymmetric vortex, for $C=2,\protect%
\gamma =0.95$, and $k=4$ ($P=66.83$). This unstable mode blows up in its
active component, as is common for $\mathcal{PT}$-symmetric systems. }
\label{instability}
\end{figure}

A unique dynamical property of the symmetric OnVs is that, when they are
unstable, they \emph{do not blow up}, unlike the FSs, and unlike
antisymmetric OnVs, which do blow up if being unstable, as shown in Fig. \ref%
{instability}(b). Instead, as seen in Fig. \ref{instability}(a), the
instability transforms symmetric vortices into \emph{stable} symmetric FSs.
Thus, the symmetric vortices demonstrate remarkable robustness as
self-trapped modes of the $\mathcal{PT}$-symmetric 2D nonlinear system.

\begin{figure}[tbp]
\centering{\label{fig8a} \includegraphics[scale=0.32]{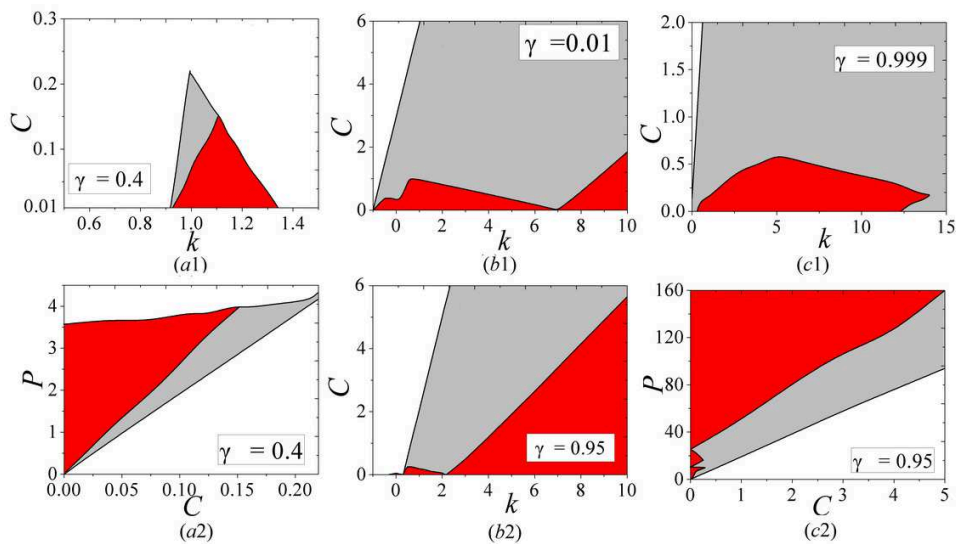}}
\caption{Stability and existence regions for $\mathcal{PT}$-symmetric vortex
solitons: (a1) in the $\left( k,C\right) $ plane, and (a2) in the $(P,C)$
plane, for a fixed gain-loss coefficient, $\protect\gamma =0.4$. The
variation of $\protect\gamma $ produces small changes in these diagrams (not
shown here in detail). (b1,b2,c1) Stability regions of $\mathcal{PT}$%
-antisymmetric vortex solitons in the $\left( k,C\right) $ plane: (a) for $%
\protect\gamma =0.01$, (b) for $\protect\gamma =0.95$, and (c1) for $\protect%
\gamma =0.999$. (c2) The stability region in the $(P,C)$ plane for $\protect%
\gamma =0.95$. In this case, unlike what happens with symmetric vortices,
the stability region strongly changes with the variation of $\protect\gamma $%
. }
\label{StaSONVS}
\end{figure}

The most important results for the symmetric and antisymmetric vortices,
namely, their existence and stability areas in planes of control parameters $%
\left( k,C\right) $ and $\left( C,P\right) $, are summarized in Fig. \ref%
{StaSONVS}. For the symmetric OnVs, it is not excluded that, above the
triangular existence regions in Fig. \ref{StaSONVS}(a1), and above the
existence region in Fig. \ref{StaSONVS}(a2), there exist very strongly
unstable vortices which could not be produced by the numerical method.

Numerical data demonstrate that dependence $P(k)$ for the symmetric and
antisymmetric vortices alike is almost exactly given by
\begin{equation}
P_{\mathrm{OnV}}=4P_{\mathrm{FS}}(k),  \label{4}
\end{equation}%
where $P_{\mathrm{FS}}(k)$ is the respective dependence for the fundamental
solitons, see Eq. (\ref{P(k)}). This relation is naturally explained by the
fact that the on-site-centered vortex may be constructed, essentially, as a
set of four weakly interacting FSs, with phase shifts $\pi /2$ between
adjacent ones \cite{Malomed}. Relation (\ref{4}) also helps to map the
existence and stability regions for the OnVs in the $\left( C,P\right) $
parameter plane in Figs. \ref{StaSONVS}(a2) and (c2).

Actually, the fact that the symmetric OnVs (at least those which could be
produced by the numerical method) exist only at relatively small values of $P
$ [at $P<4$ in Fig. \ref{StaSONVS}(a2)] explains that, as stressed above,
their instability leads not to the blow-up, but rather to the spontaneous
transformation into stable symmetric FSs. It is worthy to note too that the
stability region for the symmetric vortices in Fig. \ref{StaSONVS}(a2) is
located above the instability area (at larger $P$), while for the symmetric
FSs the situation was opposite, cf. Figs. \ref{SAFS}(c1) and 3(c2). In this
respect, the symmetric vortices resemble \emph{antisymmetric} FSs, as
suggested by the comparison with Fig. \ref{SAFANS}(d). However, the
symmetric OnVs share with symmetric FSs the trend to exist and be stable at
values of the total power essentially lower that those at which
antisymmetric counterparts are stable, hence the symmetric vortices are more
appropriate for the experimental creation.

The stability areas for the antisymmetric OnVs, displayed in Figs. \ref%
{StaSONVS}(b1)--\ref{StaSONVS}(c2), resemble their counterparts shown in
Fig. \ref{SAFANS} for antisymmetric FSs. In particular, the stability zone
tends to extend to $k\rightarrow \infty $, i.e., $P\rightarrow \infty $,
except for the case when the gain-loss coefficient, $\gamma $, is very close
to its critical value, $1$, see Fig. \ref{StaSONVS}(c1) for $\gamma =0.999$.

\begin{figure}[tbp]
\centering{\label{fig10a} \includegraphics[scale=0.37]{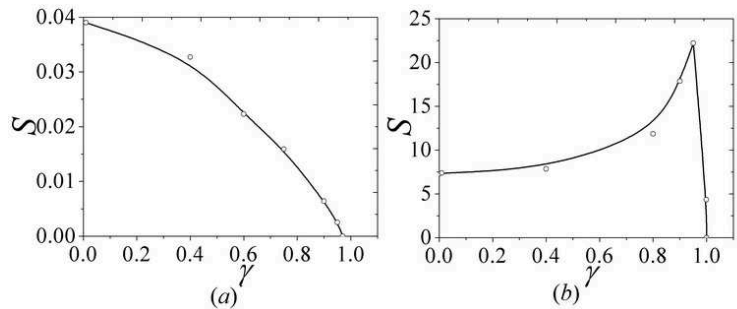}}
\caption{The dependence of the total area of the stability regions in the $%
\left( k,C\right) $ parameter plane on the gain-loss coefficient, $\protect%
\gamma $, for the symmetric (a) and antisymmetric (b) vortex solitons. The
continuous lines are only guides to the eye.}
\label{Sarea}
\end{figure}

The overall dependence of the stability of the vortices on the gain-loss
parameter $\gamma $ is characterized by total areas of the stability regions
in the $\left( k,C\right) $ plane [see Figs. \ref{StaSONVS}(a1) and
8(b1)--(c1)], which are shown as functions of $\gamma $ in Fig. \ref{Sarea}.
The drastic difference between these dependences for the symmetric and
antisymmetric vortices is obvious.

It was argued above that the fundamental $\mathcal{PT}$-symmetric FS are
likely candidates to the role of the system's ground state. To further
verify this conjecture, in Table 1, we compare values of the chemical
potential, $\mu $, for a typical set of the control parameters, $(C,\gamma
)=(0.05,0.1)$ and $P=1.9$. All the species of discrete solitons considered
above exist and are stable at this point (symmetric and antisymmetric FSs
and OnVs). The table demonstrates that the symmetric FS has a deep minimum
of $\mu $, thus definitely representing the ground state, the three other
species being excited states. It is worthy to note too that the lowest
excited state is the $\mathcal{PT}$-symmetric vortex, but not the
antisymmetric FS. Similar relations between chemical potentials of the four
soliton species have been found at other values of the parameters.

\begin{table}[tbp]
\caption{Chemical potentials of the four coexisting soliton species at $(C,%
\protect\gamma )=(0.05,0.1)$ and $P=1.9$ }\centering{\
\begin{tabular}{lccr}
\hline
Types of solitons & $\mu$ &  &  \\ \hline
symmetric fundamental soliton & -1.845 &  &  \\
symmetric on-site vortex & -1.133 &  &  \\
antisymmetric fundamental soliton & 0.145 &  &  \\
anti-symmetric on-site vortex & 0.858 &  &  \\ \hline
\end{tabular}%
}
\label{tab:table1}
\end{table}

\section{Conclusion}

The objective of this work is to introduce the 2D network of $\mathcal{PT}$%
-symmetric dimers with the cubic nonlinearity, which support stable $%
\mathcal{PT}$-symmetric and antisymmetric fundamental and vortical discrete
solitons. While the shape of these solitons can be reduced to that of their
counterparts in the conservative two-component system (although discrete
vortices were not systematically studied even in the conservative system
with two linearly coupled components), the study of the stability is
crucially important for the consideration of the $\mathcal{PT}$-symmetric
system. It has been found that the symmetric FS (fundamental soliton)
represents the ground state of the system, and may be stable at lowest
values of the total power, thus being the most promising target for the
experimental realization. On the contrary, anti-symmetric FS and OnV
(on-site-centered vortex) tend to be stable at higher powers. The symmetric
OnV represents the first metastable mode above the FS ground state. It also
tends to be stable at relatively low powers, and demonstrates remarkable
robustness: unlike all the other discrete-soliton species found in this
system, unstable symmetric OnVs do not suffer the blow-up, but rather
spontaneously transform into stable FSs.

The above analysis did not address mobility of the solitons, but, unlike the
1D version of the system, mobility of 2D discrete solitons is not plausible
in the case of the cubic nonlinearity \cite{Johansson,Johansson2,Panos}. In
this connection, it may be interesting to study 2D networks of\ $\mathcal{PT}
$-symmetric dimers with the quadratic nonlinearity, cf. Refs.\cite{chi2-1,chi2-2,chi2-3}, and also Ref. \cite{Susanto}, where the mobility was
demonstrated for 2D discrete solitons in the conservative systems with the
quadratic nonlinearity. Another promising extension may deal with 2D systems
built of elements which include nonlinear $\mathcal{PT}$-symmetric gain and
loss terms \cite{Yaro,Miro}.

\begin{acknowledgments}

This work is supported by the National Natural Science Foundation of China
(Grant Nos.11104083, 11204089 and 11205063). B.A.M. appreciates a visitor's grant for
outstanding overseas scholars (China), received via the East China Normal
University.

\end{acknowledgments}

%

\bibliographystyle{plain}
\bibliography{apssamp}

\end{document}